\documentstyle[rotate,epsf]{l-aa}
\begin{document}
\title {Global three-dimensional simulations of magnetic field
 evolution in a galactic disk}
\subtitle{I. Barred galaxies}
 \author{K.~Otmianowska-Mazur\inst{1,4}, S.~von Linden\inst{2},
         H.~Lesch\inst{3}, G.~Skupniewicz\inst{1}}
 \offprints{ K.~Otmianowska-Mazur (Krak\'ow), e-mail: otmian@oa.uj.edu.pl}
 \institute{ Astronomical Observatory, Jagiellonian University,
     ul. Orla 171, Krak\'ow, Poland
 \and Landessternwarte Heidelberg, K\"onigstuhl, D-69117 Heidelberg, Germany.
 \and Universit\"ats-Sternwarte, Scheinerstr. 1, D-81679 M\"unchen, Germany
 \and Max-Planck Institut f\"ur Radioastronomie, 
 Auf dem H\"ugel 69, 53121 Bonn, Germany}

 \thesaurus{02.13.2; 11.13.2; 11.11.1; 11.19.2}
 \date{Received 28 June 1996}
 \maketitle
 \markboth{K.~Otmianowska-Mazur et al.: On 3D galactic magnetic 
 field evolution}{}

%--------------------------------------------------------------------
\begin{abstract}

The evolution of three-dimensional, large-scale, magnetic fields, 
in a galactic disk is investigated numerically.
The N-body simulations of galactic dynamics are incorporated
into the kinematic calculations of induction equations to study
the influence of non-axisymmetric gas flows on the galactic magnetic
field. The time-dependent gas velocity fields are introduced as
input parameters for the MHD-simulations. The effects of interstellar
turbulence are given as the turbulent diffusion of the magnetic field.
Our principal concern  is to check how dynamical evolution of the 
galactic gas affects the global magnetic field structure and intensity.

We have found that the magnetic field responds sensitively to
changes in the gas velocity field, and even slight variations
of the dynamical parameters, such as the gas mass/total mass ratio results in
nonuniform intensity structures, i.e. magnetic arms.  The magnetic 
lines of force are well aligned with spiral arms and bar due 
to compressional flows in these features. In the inter-arm regions the
areas with magnetic vectors going opposite to the main magnetic spirals
are present. Moreover we could identify Parker-like intensity features
perpendicular to the disk plane. Obviously even
weak perturbations of the velocity field already produce almost instantaneous
changes in the local magnetic field geometry. This proves the importance of
our approach to use the velocity fields resulting from global dynamical
instabilities for magnetohydrodynamical simulations.

\end{abstract}
\keywords{{MHD; - 
galaxies: magnetic fields; kinematics and dynamics; spiral}

\section{Introduction}

Spiral galaxies are magnetised disks. The evolution and origin of
their large-scale magnetic field is supposed to be related to induction
processes, via motion of highly conducting plasmas (for recent reviews
see Wielebinski and Krause 1993; Kronberg 1994; Beck et al. 1996).
This relation between the plasma motion and the magnetic field is well
known as a dynamo. The conventional dynamo theory contains a 
large-scale velocity
typically given by the differential rotation of the disk plus some
turbulent contribution which describes the turbulent diffusion and the helicity.
Since the helicity is of the order of $1\,$ km s$^{-1}$, it is not an 
observable quantity
and its value and physical origin can only be determined from 
theoretical considerations
(Hanasz and Lesch 1993; Beck et al. 1996).

The dominant term for magnetic field amplification is the large-scale
velocity field, which is a really observed quantity. The large-scale
velocity field of a spiral galaxy is the product of the differentially
rotating disk, an almost rigidly rotating galactic bulge, and non-axisymmetric
velocity components, which are the result of dynamical instabilities
(e.g.  Binney and Tremaine 1987
and references therein). Whereas the energy density of the
large-scale magnetic field is a negligible portion of the rotational
energy density of a galactic disk ($W_{\rm mag}\propto 10^{-13}\,
{\rm erg\, cm}^{-3}$
($B\sim 5\mu$G), $W_{\rm rot}\propto 10^{-10}\, {\rm erg\, cm}^{-3}$
($v_{\rm rot}\sim 200\, {\rm km\, s}^{-1}$)),
it is comparable with the kinetic energy density of the 
perturbations driven by dynamical instabilities
like spiral arms and bars ($W_{\rm dyn}\sim 10^{-12}\, {\rm erg\, cm}^{-3}$
($v_{dyn}\propto 10\, {\rm km\, s}^{-1}$)).
Since the dynamical instabilities present global responses of the
self-gravitating, dissipative component of the 
interstellar medium in disk galaxies,
they will have a considerable influence on the structure and evolution
of large-scale magnetic fields in disk galaxies (Chiba and Lesch 1994).

A tight relation between the galactic velocity structure 
and the $B$-field is indeed clearly visible in
many radio observations:
recent high frequency (Faraday rotation  negligible)  observations
of radio polarisation from nearby galaxies: M83 (Neininger  et
al. 1991), M51 (Neininger 1992, Neininger and  Horellou
1996) and NGC\,6946 (Ehle \&  Beck  1993)  show  that  the  large-scale
galactic magnetic field is aligned  with the spiral  arms  and  bars,
which  suggests  the  presence  of  the  dynamical  influence   of
non-axisymmetric gas flows on the  galactic  magnetic  field.
High angular resolution  interferometer  observations  support  this
hypothesis, showing  that  the  galactic  magnetic  fields  follow
the dust lanes in M31 (Beck et al. 1989) - regions with  a
high ratio of gas compression. Narrow lines of highly  polarised
emission have been found at the inner edges of the spiral arms  in
IC 342 (Krause 1993). These facts  could  mean  a  close  dynamical
connection between magnetic fields and interstellar  gas  flows  in
the spiral arms and bars.
Lower spectral resolution radio observations, which are  more  sensitive
to extended regions of polarised emission, show  the  presence  of
such areas also  in  the  inter-arm  regions (Beck \& Hoernes 1996).
The  magnetic  field
vectors there show a spiral structure which has the same  pitch
angle as do the spiral arms in NGC\,1566 (Ehle  et  al.  1996),
M81 (Krause et al. 1989) and NGC\,4254 (Soida et
al 1996).

The appearance of the dynamical instabilities does not depend
on an existing magnetic field (Binney and Tremaine 1987). Thus, we have to
investigate the magnetic response 
to a large-scale non-axisymmetric instability.
Since the dynamical instabilities are inherently
time-dependent, the magnetic field
will also vary with time.
We will present time-dependent 3D dynamical simulations
of galactic disks, containing stars
and molecular-gas clouds. The excited instabilities will
lead to a time dependent
velocity field, which will be inserted into the induction equation.

It is the aim of this contribution to 
present and discuss the results of this procedure.
We consider
the following problems:
(1) What is the structure and evolution of  the  large-scale  galactic
magnetic field under the influence of  spiral  and  bar  structures
present in a galactic disk?
(2) To what extent could the resulting  magnetic  field  explain  the
observed spiral pattern of magnetic fields in nearby galaxies?

With the same motivation Otmianowska \& Chiba
(1995, Paper I) have
shown from their simulation that the magnetic field follows closely
the interstellar gas motions, resulting in structures aligned with
the spiral arms and the bar, a configuration 
which is in good agreement with observation
(Ehle \& Beck 1993, Neininger 1992, Neininger et al. 1991).
However, this effect is visible only near the spiral
arms and rather than in the inter-arm regions, where the derived magnetic
field vectors go outward from the galactic center. This is not
observed in actual galaxies. This fact is due to certain
simplifications
of the first model: the given gravitational potential of stars and the
single (not evolutionary) velocity field of the interstellar gas.
The present project constitutes fully dynamical evolution of the gas
velocity field over $10^{9}$ yr in 3D
and simulates two components of
interstellar medium: stars and molecular gas.
The mean-field dynamo mechanism
is not considered in our calculations.

The outlay of the paper is as follows: Sect.~2 is devoted to a description
 of the simulation methods.
The input model parameters are presented in Sect.~3, and the
results of the simulations appear in Sect.~4. Conclusions and
discussion are given in Sect.~5.

\section {Models and numerical methods}

First, to obtain realistic velocity fields for calculations of
magnetic field evolution (solving the induction equation), we perform
a three-dimensional N-body code,
using the particle-mesh scheme. Next the
resulting
velocity fields of the gaseous component are used to calculate
the three-dimensional
large-scale galactic  magnetic
field. The N-body simulations obtain two components: stars and
molecular clouds. The clouds represent the major mass component
of the interstellar gas.
Since typically the gas mass is only 10\%
of the total mass, the gas
reacts like a test particle to the gravitational potential built up by the
stellar and dark matter background.  The importance of the gas
is emphasized given by the fact that only in the
presence of the dissipative gas component are dynamical instabilities
long-lived features in a disk (Binney and Tremaine 1987). Without
a considerable amount of gas, like in S0-galaxies, no 
spiral structure or bars
appear; this is supported by many observations
and N-body simulations (Sellwood and Carlberg 1984; Kennicutt 1989).
The magnetic field is  indeed connected  with the gaseous component
of a galaxy and the instabilities responsible for structure
will produce a non-axisymmetric
velocity field, which in turn will have a significant 
influence on the magnetic field.
\subsection{N-body numerical scheme}

We use self-consistent 3D simulations for the modelling of the galaxy.
The code is a 3D version of the 2D code of Combes \& Gerin (1985)
kindly supplied by F. Combes.

The disk of the galaxy consists of stars ($N_{\rm s}\sim 38\,000$)
and clouds ($N_{\rm g} \leq 19\,000$).
The particles are interacting gravitationally with each other and
are embedded in a bulge and a halo.
The gravitational potential is calculated from the 
density via fast Fourier transforms (FFT)
using a 128 x 128 x 64 grid (with a useful grid of 64 x 64 x 32).
We calculate the density at each
spatial point and at each integration step.

Collisions between molecular clouds are highly inelastic, and a particle
model can provide the best control of the corresponding energy dissipation.
To simulate a collision in the cloud ensemble, we adopt the same
scheme and local processes as described by Casoli \& Combes (1982)
and Combes \& Gerin (1985). The cloud ensemble is represented by its
mass spectrum (masses between $5\cdot10^2$ and $5\cdot 10^5$ M$_{\odot}$).
We also take  into account that
giant molecular clouds (GMC) have a finite lifetime. After about $10^7$ years,
the clouds are destroyed by tidal
forces and/or star formation (stellar winds, supernova explosions etc.).
To describe
this destruction numerically we explicitly focus on
the GMC (mass of one GMC $M_{\rm GMC} >
2 \cdot 10^5 {\rm M}_\odot$).
After $4 \cdot 10^{7}$ years, cloudlets are reinjected into
interstellar space with a steep power-law spectrum and a 2-dimensional
random-velocity distribution. A star particle is added at this location
to the stellar disk.

We use two analytical components to simulate the bulge and
dark halo. These components are
Plummer spheres $\Phi_{\rm bulge} (r)=
-G M_{\mbox b}\cdot(r_d^2+b^2)^{-1/2}$ and
$\Phi_{\rm halo} (r)= -G M_{h} \cdot (r_h^2+a^2)^{-1/2}$.
The halo mass within disk radius $r_d$ is $M_h$, $a$ is its scale length
and the bulge mass is $M_b$ (within radius $r_h$) with a scale length $b$.

The velocity distribution of the particles is then chosen to
stabilise the disk with respect to axisymmetric disturbances.
For a stellar disk to be stable against these disturbances, the 
Toomre parameter
$Q\sim {\kappa v \over{G \Sigma}}$ has to be larger than 1.
Here $v$ denotes the velocity dispersion of the different
components (gas or stars), and $\Sigma$ is their surface mass density.
$\kappa$ is the epicyclic frequency. When $Q$ is in
the range $ 1 < Q < 4$ , the stellar disk is, however, unstable with respect
to non-axisymmetric perturbations and in particular to bar formation
(Hohl 1971).

One can define the Toomre parameter Q (Toomre 1964) separately for gas and
stars:
   \begin{equation}
Q_{\rm gas}={v_{\rm gas}\kappa\over 3.36 G \Sigma_{\rm gas}}\qquad
 Q_{\rm star}={v_{\rm star}\kappa\over \pi G \Sigma_{\rm star}}
   \end{equation}
\noindent but these are only indicative, since there is
no independent stability criterion. The coupling between gas and stars
favours instabilities: both components could be unstable
together in situations
where they would be stable separately (Jog \& Solomon 1984; Jog 1992).
The initial circular velocity is computed with the N-body integrator,
in order to take into account softening and disk truncation; then the initial
rotational velocity is computed from the Jeans equations, taking into
account the velocity dispersions.

The present code
has been tested in many simulations
(e.g. Combes et al. 1990). The collision parameters (number of particles,
elasticity of collisions, etc.)
have been varied around the values chosen here, without any effect on
the results (e.g. von Linden et al. 1996).

\subsection{Solving the induction equation}

The evolution of the large-scale magnetic field is simulated with  the
three-dimensional ZEUS code,
in which the recently developed constrained transport (CT) algorithm
is implemented for the numerical evolution of the components  of
the magnetic field in  three  dimensions  (Stone \& Norman 1992).
We  use  only  the  part of the magnetic field
evolution, where the flux-freezing approximation applies:
%%%
\begin{equation}
\partial \vec{B}/\partial t=\hbox{rot}(\vec{v}\times \vec{B}),
\end{equation}
%%%
\noindent where $\vec{v}$ is the velocity of gas and $\vec{B}$ is the
magnetic induction.
The numerical method fulfills automatically the  condition
$\hbox{div}{B}=0$. Diffusion effects are approximated by
"pulsed-flow" method (Bayly \&  Childress 1989), where first
evolution of perfectly frozen  in  magnetic  field is considered. Then
after a few tens of time steps the effects of the  magnetic diffusion are
introduced using the solution of the diffusion equation  with
an application of the Green's function (Otmianowska-Mazur et  al.  1992).
In practice this is equivalent to convolving  the  whole grid of
magnetic field components with a Gaussian function having
dispersion  $\sqrt{(2\eta \Delta t)}$, where $\eta$ is the magnetic
field diffusion coefficient.
The convolution method also satisfies the divergence-free
condition of the magnetic field.
Computations with the solution of the induction
equation, including  magnetic diffusion,
are in preparation.

To incorporate the particle-based flow into the grid-based code,
we adopt the approximation of a spline function.
The method is described in detail in Paper I
(Paper I, eq. no.4 \& ref. therein).

The computations have been performed using a Cray Y-MP in HLRZ J\"ulich
and Convex Exemplar SPP-1000 Series in ACK CYFRONET-KRAK\'OW.

\section{Input models parameter}

\subsection{N-body model}

We simulate non-interacting disk galaxies using a 3D code
to study the effect on the vertical extension of the disk.

The time step of the N-body simulation is 10$^6$ yr. The
time step for collisions has not to be so frequent, since the
collision time-scale is of the order of  10$^7$ yr.

The stellar particles are initially distributed in a Toomre disk (Toomre 1963)
of mass $M_{\mbox d}$ within disk radius $r_{\mbox d}$
and scale length $d$.
The gas clouds are distributed in an exponential disk, with scale-length
12 kpc, and mass $4 \cdot 10^9$ M$_{\odot}$,
i.e. corresponding to 4.4 \% of the disk mass.

We made several different simulations by varying the input parameters
for the disk and the bulge. Depending on the input parameters, the
simulations produce bar and/or spiral structures or
only axi-symmetric structure like rings.
We are primary interested in the qualitative results. Thus
we present here only barred galaxies
and we focus on the development
of strong bars. Friedli \& Benz (1993, 1995) explore
the influence of different initial conditions
of the gas, disk, bulge, and halo mass on the formation of
bars. For this reason, we concentrate on two basic experiments which
differ in the input parameters (see Table~1).

\begin{table}
\begin{flushleft}
\begin{tabular}{llllll}
\hline\noalign{\smallskip}
% model gal8 and model gal14
Model    &I      & II\\
\hline\noalign{\smallskip}
Mass in $10^{10} {\rm M}_\odot$:     &     &  \\
- dark-halo mass $M_h$               & 9.6 & 9.6  \\
- disk mass    $M_d$                 & 7.2 & 7.7  \\
- gas mass  $M_g$                    & 0.4 & 0.4  \\
- bulge mass   $M_b$                 & 4.8 & 4.3\\
- mass ratio $(M_d+M_g)/M_{\rm tot}$ & 0.35& 0.37 \\
Scale length in kpc: & & \\
- dark halo $h$      & 15 & 15\\
- bulge      $b$     &1.85&2\\
- disk  $d$          &6&6\\
\noalign{\smallskip}
\hline
\end{tabular}
\caption[]{Input parameters for model I and II}
\end{flushleft}
\end{table}

\subsection{Magnetic field evolution model}

The model parameters adopted in the computations are summarised
in Table 2. Our experiments are divided into two main groups:
magnetic field evolution in a
galaxy possessing gravitationally strong (I) and weak (II) bar disturbance
(Table~2).

The kinematic evolution of the magnetic field is solved
in a 3D rectangular grid of points, where the XY plane is the galactic
plane and the Z-axis is the axis of galactic rotation pointing
in the northern direction. The rectangular 
size in the  X and Y directions is 30 kpc
with the grid interval
of 600 pc  (51 grid points) or 300 pc (101 grid points), depending
on the experiment. The height of the model galaxy
is 1 kpc with spatial resolution of 50 pc (21 grid points) (Table~2).
The adopted time
step is $2\cdot 10^{5}$ yr. Computations with a smaller time step,
$10^{4}$ yr, give identical results.
The input magnetic field configuration is purely axisymmetric, going in
the anti-clockwise direction,
although a more general structure would be the 
superposition of several magnetic
modes (see Paper I for discussion).
The initial field is toroidal, its intensity decreases toward the
galaxy center (from $R=4$ kpc) to avoid numerical instabilities 
and at radii larger than 4 kpc is taken as a constant. At $R>12$ kpc
it decays exponentially. The z-component is initially zero.
A calculation with more realistic seed fields is in
preparation. The intensity of the initial field is
$10^{-6}$ G, since fields of this order are
widely observed in spiral galaxies and the
actual value of the initial field does not change the results (Paper 1).
We adopt so called
"outflow" boundary conditions (Stone \& Norman, 1992).
Physically this means
that magnetic flux outflow from the galaxy is present.
Such boundary conditions result in a strong diminishing of
the magnetic energy density during all the experiments. The
decrease of magnetic energy is
connected with the effective magnetic Reynolds number in the ZEUS code.
The magnetic field decay simulations made by Hawley et al. (1996) show that
the diffusivity provided only by numerical truncation errors gives
Reynolds numbers of order $10^{3}$ and $10^{4}$, depending
on the grid spacing (Hawley et al., 1996). Our concern
is to clarify how the numerical diffusion affects our magnetic simulations.
To solve this problem, the calculations for model I
are performed for two cases (see Table~2):
51 grid points in the X and Y directions with step size
of $600$ pc (Ia) and 101 grid points with spacing $300$ pc (Ib).
Both experiments are made with no additional diffusion ($\eta=0$)
to check how the grid spacing effects the truncation.
  \begin{figure}[tbh]\rule{0.0pt}{1cm}
   \epsfysize=8.5cm
    \rotate[r]{\epsffile[90 175 520 623]{5461F1a.ps}}
\caption[]{\label{gal8}{\bf a:}  Model I at time step $8 \cdot 10^8$ yr.
    The upper picture shows the total gas (right)
    and 9500 stellar (left) particles of the $x-y$ plane.
    The lower picture shows the belonging $y-z$ plane of the simulation.}
\end{figure}
\addtocounter{figure}{-1}
  \begin{figure}[tbh]\rule{0.0pt}{1cm}
   \epsfysize=8.5cm
    \rotate[r]{\epsffile[90 175 520 623]{5461F1b.ps}}
\caption[]{{\bf b:} Model I at
$8.5\cdot 10^{8}$ yr}
\end{figure}\addtocounter{figure}{-1}
  \begin{figure}[thb]\rule{0.0pt}{1cm}
   \epsfysize=8.5cm
    \rotate[r]{\epsffile[90 170 520 623]{5461F1c.ps}}
\caption[]{{\bf c:} Model I at $9\cdot 10^{8}$ yr}
\end{figure}\addtocounter{figure}{-1}
  \begin{figure}[bth]\rule{0.0pt}{1cm}
   \epsfysize=8.5cm
    \rotate[r]{\epsffile[90 170 520 623]{5461F1d.ps}}
\caption[]{{\bf d:} Model I at $10^{9}$ yr}
\end{figure}

We adopt a diffusion coefficient $\eta$ based on the
turbulent motions of the interstellar gas.
The value of $\eta$ changes from $10^{26}~{\rm cm}^{2}$/s (models
Ic and IIb) by $3.6\cdot
10^{26}$ cm$^{2}$/s (Ie) to
$10^{27}$ cm$^s$/s (Id) and is two or three-dimensional
(see the discussion below).
The computations
can last no longer than about $6\cdot 10^{8}$ yr, because of the
significant decrease of magnetic energy due to numerical dissipation and
the outflow boundary conditions.
In order to check late as well as early
epochs of the velocity field evolution
the calculation are performed from $7\cdot 10^{8}$ to about $10^{9}$ yr
for model I
and from  $2\cdot 10^{8}$ to about $8\cdot 10^{8}$ yr for model II.
The computations have also been done for two dimensional
diffusion working only in the XY plane
(Table~2, Id and Ie) to study the possibility
of much smaller magnetic dissipation in the Z
direction than in the galactic plane.
The diffusion coefficient value in this case
is $10^{27}$ cm$^2$/s (Id) or $3.6\cdot 10^{26}$ cm$^2$/s (Ie).

\begin{table}
\begin{flushleft}
\begin{tabular}{lllllll} \hline\noalign{\smallskip}
Model & $dx(=dy)$ & $dz$ & $N_{\rm x}\cdot N_{\rm y}\cdot N_{\rm z}$
& $\eta$ & \\
      & pc   & pc&     & cm$^2$/s&\\
\noalign{\smallskip}\hline\noalign{\smallskip}
Ia    & 600 & 50& $51\cdot 51\cdot 21$& 0& \\
Ib   & 300 & 50& $101\cdot 101\cdot 21$& $0$& \\
Ic   & 600 & 50& $51\cdot 51\cdot 21$& $10^{26}$& \\
Id   & 600 & 50& $51\cdot 51\cdot 21$& $10^{27}$planar& \\
Ie   & 300 & 50& $101\cdot 101\cdot 21$& $3.6 \cdot 10^{26}$planar& \\
\noalign{\smallskip}\hline\noalign{\smallskip}
IIa   & 600 & 50& $51\cdot 51\cdot 21$& $0$& \\
IIb   & 600 & 50& $51\cdot 51\cdot 21$& $10^{26}$& \\
\noalign{\smallskip}\hline
\end{tabular}
\end{flushleft}
\caption[]{The magnetic field evolution model parameters}
\end{table}

The choice of two diffusive models needs some more explanation. In our
philosophy, the magnetic field evolution is a direct consequence of the
dynamical instabilities driven by non-axisymmetric perturbations of the
gravitational potential. 
The molecular clouds are very sensitive to such perturbations.
The formation of the clouds also needs an instability, which produces the
local gravitational collapse. In disk systems, such a local process
can be described by a global stability criterion, 
depending on the disk rotation,
i.e. the epicyclic frequency and the surface mass 
density $\Sigma$ (Binney and Tremaine
1987). In terms of the Q-value (see Eq. 1) 
the condition for axisymmetric collapse can be rewritten as
$Q<1$ and a typical collapse scale $\lambda_{\rm c}$ and velocity dispersion
$v_{\rm c}$ can be defined with the critical value $Q=1$ (Lesch 1993)

\begin{equation}
\lambda_{\rm c}={2\pi^2 G\Sigma\over{\kappa}}
\qquad {\rm and}
\qquad
v_{\rm c}={\pi G\Sigma\over{\kappa}}.
\end{equation}
As an initial condition we assume that the disk already contains
molecular clouds. That is, the disk is kept on the border of gravitational
instability, creating a cloudy medium with both cloud sizes and 
separation of order $\lambda_{\rm c}$.
These clouds should provide the angular momentum transport by a two dimensional
random walk with a step length $\lambda_{c}$ and velocity $v_{c}$, i.e. the
viscosity is (Lin and Pringle 1987)

\begin{equation}
\nu\sim \lambda_{\rm c}v_{\rm c}\sim \lambda_{\rm c}^2 \Omega(R).
\end{equation}

We note that from the onset condition $Q<1$, one can define a critical
surface mass density $\Sigma_{\rm c}$. If the theory is correct 
then for $\Sigma>\Sigma_{\rm c}$,
cloud formation and star formation should start. Indeed, 
Kennicutt (1989) and Elmegreen et al. (1994)
found that in disk galaxies of all Hubble types, 
the size of star forming regions
is of the order of $\lambda_{\rm c}$, and that the star formation
starts at a radius where
$\Sigma\sim \Sigma_{\rm c}$. They concluded that the star forming threshold is
associated with the onset of the large-scale 
gravitational instability in the galactic gas disks.

Thus, the gravitational disk-instability 
may serve as the ultimate pool  for turbulence
in the interstellar medium. We take the gravitational viscosity equivalent
to the magnetic diffusivity.
According to Eq.\,4  a typical galactic disk model gives a characteristic
value of
\begin{equation}
\nu\sim 10^{26}\, cm^2 s^{-1} \left[ {\lambda\over 100pc} \right]^2 \left[
{\Omega \over 10^{-15} s^{-1}}\right]
\end{equation}
(Lesch 1993). Since the cloud formation is
restricted to the disk, the diffusivity acts only radially in the
disk plane, provided the disk activity is low, so that no considerable gas flow
into the halo is expected. This is obviously the case 
in most disk galaxies, which
show no extended radio halos (Dumke et al. 1995).

\begin{figure}[t]\rule{0.0pt}{1cm}
   \epsfysize=8.5cm
    \rotate[r]{\epsffile[90 170 520 623]{5461F2a.ps}}
\caption[]{\label{gal14}{\bf a:}  Simulation for model II at
    time step $2 \cdot 10^8$ yr (see caption Fig.\,1a).}
\end{figure}
\addtocounter{figure}{-1}
  \begin{figure}[t]\rule{0.0pt}{1cm}
   \epsfysize=8.5cm
    \rotate[r]{\epsffile[90 170 520 623]{5461F2b.ps}}
\caption[]{{\bf b:} Model II at $3\cdot 10^{8}$ yr}
\end{figure}\addtocounter{figure}{-1}
  \begin{figure}[t]\rule{0.0pt}{1cm}
   \epsfysize=8.5cm
    \rotate[r]{\epsffile[90 170 520 623]{5461F2c.ps}}
\caption[]{{\bf c:} Model II at $5\cdot 10^{8}$ yr}
\end{figure}

\section{Results}

\subsection{Model I}%gal8
During the first time steps the disk exhibits a strong spiral
structure.

Since the stars are collision-free the stellar spiral arms are
much broader than the gaseous spirals.
The winding of spiral arms can lead to the formation of pseudo-ring
structures (Friedli \& Benz 1993) which is seen in the very outer part
of the disk.
%%%%%%%%%%%%%%%%%%%%%%%%%%%%%%%%%%%%%%%%%%%%%%%%%%%%%%%%%%%%%%%%%%
%%%% Fig.3a %%%%%%%%%%%%%%%%%%%%%%%%%%%%%%%%%%%%%%%%%%%%%%%%%%%%%%%
%%%%%%%%%%%%%%%%%%%%%%%%%%%%%%%%%%%%%%%%%%%%%%%%%%%%%%%%%%%%%%%%%%

  \begin{figure*}[tbh]\rule{0.0pt}{1cm}
   \epsfysize=15.5cm
    \epsffile[60 30 340 493]{5461F3a.ps}
\caption[]{\label{fig3}{\bf a:}
 Magnetic field vectors (left) and
 intensity of magnetic field (gray plot, right) in the galactic plane
 for the model Ib with no
 magnetic diffusion at $t~=~8\cdot 10^{8}~$yr (top) and at $t~=~8.5\cdot
 10^{8}~$yr
 (bottom)}
\end{figure*}
%%%%%%%%%%%%%%%%%%%%%%%%%%%%%%%%%%%%%%%%%%%%%%%%%%%%%%%%%%%%%%%%%%
%%%% Fig.3b %%%%%%%%%%%%%%%%%%%%%%%%%%%%%%%%%%%%%%%%%%%%%%%%%%%%%%%
%%%%%%%%%%%%%%%%%%%%%%%%%%%%%%%%%%%%%%%%%%%%%%%%%%%%%%%%%%%%%%%%%%
\addtocounter{figure}{-1}
\begin{figure*}[hbtp]
\epsfysize=15.5 cm
\epsffile[60 40 340 493]{5461F3b.ps}
\caption[]{{\bf b:}
 Magnetic field vectors (left) and
 intensity of magnetic field (gray plot, right) in the galactic plane
 for the model Ib with no
 magnetic diffusion at $t~=~9\cdot 10^{8}~$yr (top) and at
 $t~=~9.5 \cdot 10^{8}~$yr
 (bottom)}
\end{figure*}

After $3\cdot 10^8$ yr a 3-armed spiral structure appears.
The mass transport, initiated by the spiral arms, leads
to a strong mass concentration in the central region, which
changes the behaviour of the disk. First, the mode
changes into a 2 armed spiral and few time steps later
the mass concentration allows the formation of a very massive and
long bar. This bar reaches its maximal strength
within a few time steps. After $8\cdot 10^8$ yr the bar is 25\,kpc long
and ended at the bar-corotation radius (12.5 kpc) (see Fig.\,1a).

The gas bar is very thin in comparison to the stellar one (see
also the simulation CSG1 from Combes \& Elmegreen 1993).
During the bar evolution a 2 armed spiral mode is excited in the outer disk
radii is starting at the bar-corotation radius.

The stars are collision-free so their Q-parameter
is high. Therefore the
stellar disk is thicker (dynamically hotter) than the gas disk, which
can be seen in the $y-z$ diagram (see Fig.\,1).
The evolution of the stellar disk is very similar to the simulation
made by Gerin et al. 1990. We found the same peanuts shape bulge in the $y-z$
diagram, which is formed due to the bar formation.

%%%%%%%%%%%%%%%%%%%%%%%%%%%%%%%%%%%%%%%%%%%%%%%%%%%%%%%%%%%%%%%%%%
%%%% Fig.4a%%%%%%%%%%%%%%%%%%%%%%%%%%%%%%%%%%%%%%%%%%%%%%%%%%%%%%%
%%%%%%%%%%%%%%%%%%%%%%%%%%%%%%%%%%%%%%%%%%%%%%%%%%%%%%%%%%%%%%%%%%
\begin{figure*}[hbtp]
\epsfysize=15.5cm
\epsffile[60 40 340 493]{5461F4a.ps}
\caption[]{{\bf a:}
	Magnetic field vectors (left) and
intensity of magnetic field (gray plot, right) in the galactic plane
for the model Id with the
2D magnetic diffusion at $t = 8.5\cdot 10^{8}~$yr (top) and at
$t = 9\cdot 10^{8}~$yr
 (bottom)}
\end{figure*}
%%%%%%%%%%%%%%%%%%%%%%%%%%%%%%%%%%%%%%%%%%%%%%%%%%%%%%%%%%%%%%%%%%
%%%% Fig.4b %%%%%%%%%%%%%%%%%%%%%%%%%%%%%%%%%%%%%%%%%%%%%%%%%%%%%%%
%%%%%%%%%%%%%%%%%%%%%%%%%%%%%%%%%%%%%%%%%%%%%%%%%%%%%%%%%%%%%%%%%%
\addtocounter{figure}{-1}
 \begin{figure*}[tbh]\rule{0.0pt}{1cm}
   \epsfysize=8.5cm
    \epsffile[50 275 427 517]{5461F4b.ps}
\caption[]{{\bf b:}
Magnetic field vectors (left) and
intensity of magnetic field (gray plot, right) in the galactic plane
for the model Id with the
2D magnetic diffusion at  $t~=~9.5 \cdot 10^{8}$~yr}
\end{figure*}
The bar pattern speed decreases slowly (Fig.\,1b and c), but
the bar is visible until the end of our simulation ($1.4\cdot 10^9$ yr)
(Fig.\,1d).
We present here the same time steps of the dynamic evolution as
we show for magnetic field experiment (see Sect. 4.3).

\subsection{Model II}%gal14
Initially this model is
similar to the case I. A spiral structure appears earlier in the
disk than in model I,
(after $2 \cdot 10^8$ yr) (see Fig.\,2a).
the arms are more open than in the first simulation (Fig.\,2b).
The mass concentration in the inner disk is higher than in model I.
 For example after $9 \cdot 10^8$ yr
model II has accumulated 30\%
more mass in the inner 1 kpc than model I.

After $4\cdot 10^8$ yr a bar is formed in the disk.
This bar increases its length during the next 10 time steps
(see Fig.\,2c).
The bar is less massive and shorter than
in the first simulation
and has his longest extend (20\,kpc) after $6\cdot 10^8$ yr.
In the outer part of the disk a two armed spiral pattern is
situated like in the simulation made by Gerin et al. 1990 .
This has been verified by an analysis of the potential
via FFT like Sparke \& Sellwood  (1985).
The ring-like structure at the bar belongs to
orbit families at Lagrange point (Contopoulous \& Papayannopoulos 1980).

In the z-direction one can see again that the gas disk in the inner part gets
thinner during this evolution. In the outer part of the disk r$>$30 kpc
the disk gets thicker (3 kpc, Fig.\,2 y-z diagram).

During the next time steps the bar shrinks and more mass is concentrated
at the bar.
We follow the bar evolution for $1\cdot 10^9$ yr.

\subsection{The 3D evolution of magnetic field: model I}

The calculations with the strong bar disturbance (model I) have been
done for five sub-cases:
Ia and Ib with no magnetic diffusion (with 51 and 101 grid points
in the X and Y axes to check the rate of numerical dissipation),
Ic with
the diffusion in three dimensions ($\eta=10^{26}$cm$^2$/s, 51 grid
points),
Id with the planar diffusion (only in the XY plane,
$\eta=10^{27}$cm$^{2}$/s, 51 grid points)
and Ie with the planar diffusion ($\eta=3.6\cdot 10^{26}$cm$^2$/s, 101
grid points). The results concerning
the evolution of a magnetic field structure are discussed,
as an example,
for the case Ib and Id. The rest of
computations for the model I are
presented in the discussion about
magnetic energy density time evolution, later in this paragraph.
In Fig.\,1a-d the distribution of the stellar and gaseous
disk for the model I is presented for the same evolutionary stages.
We can see that the whole galactic
structure is clearly visible for stars and for gas.  The  system
rotates in the anti-clockwise direction. The evolution of magnetic
field for the case Ib is presented in Fig.\,3a (top)
for  $8\cdot10^{8}~$yr,  where the
rotation of completely frozen-in magnetic field, which follows the
motions of gas is shown. The magnetic field lines are aligned with the bar
and spiral arms, however in the inner layers of both arms a small
reversals of magnetic field vectors (shown in the galactic plane) could be seen
(Fig.\,3a, top-left). The main magnetic arms are distributed slightly outside
the main molecular arms.
The magnetic intensity
distribution in the galactic plane 
(Fig.\,3a top-right, gray plot) shows that the magnetic
spirals, as well as spiral arms from molecular clouds do not form
a single, well described feature. The magnetic arms are dispersed into
two, three parts, where the direction of vectors could be opposite to
the initial field configuration. There could be also seen magnetic bridges
between two closely distributed magnetic arms (Fig.\,3a, top-right).
The magnetic field vectors start to
be mixed at the ends of the
bar (Fig.\,3a, top-left). 
For $8.5\cdot10^{8}~$yr (Fig.\,3a, bottom)
the  regions, where  field  direction  changes
abruptly, are also present.
The magnetic  field configuration begins to be
more complicated than the velocity field in the same
moment of time, what is present also in the magnetic evolution performed
in the 2D experiment (Otmianowska-Mazur et al. 1995).
The regions with magnetic vector reversal are distributed in the
inner layers of the magnetic arms (Fig.\,3a, bottom-left).
The magnetic intensity maxima
are visible again in the spiral arms and along the bar (Fig.\,3a, bottom-right)
and the reversal regions show significantly lower intensity.
For $9\cdot10^{8}$ yr (Fig.\,3b, top) the rotation of the
structure follows, the galaxy rotates once,  field  is
more and more twisted and two additional  spiral  arms,
going in the opposite direction than main spirals, form.
The magnetic strength, as it is seen in Fig.\,3b top-right (gray plot),
is highest in the dynamical structure, however regions with reversals
are here better visible than for $t~=~8.5\cdot 10^{8}$ yr,
especially for the upper magnetic arm. 
The last graph for $9.5\cdot 10^{8}~$yr presents
regions where even
three reversals of magnetic field vectors could be seen
(Fig.\,3b, bottom-left). The magnetic configuration shows
two arms in the upper part of the figure with opposite
vector direction. The intensity of the arm going
according to the input magnetic direction is certainly higher,
which is visible in the magnetic intensity gray plot (Fig.\,3b, bottom-right).
In the lower part of Fig.\,3b, bottom-left, only one
magnetic arm is visible with normally directed vectors;
the second arm shows opposite direction.
The magnetic evolution with high Reynolds
numbers shows
that the influence of a dynamical galactic evolution on a magnetic
field history is much more complicated than we expected.
The magnetic field configuration, as well as the velocity field,
rotates in the anti-clockwise direction in the whole evolutionary
time. The numerical simulations also present that the magnetic field
vectors are mixed at both ends of a bar (e.g. Fig.\,3a, bottom-left).
The analysis of gas density maxima distributions shows that maxima of
magnetic intensity are distributed slightly outside the molecular
spirals.

The presence of the magnetic diffusion changes the picture significantly
(case Id, Fig.\,4). For the presentation we choose 
the case with the highest value
of the coefficient $\eta=10^{27} \rm cm^{2}/\rm s$.
At the first evolutionary stages the magnetic
field structure is not influenced strongly by the diffusion.
The situation changes
for $8.5\cdot10^{8}$yr (Fig.\,4a, top), when the difference between
configurations with and without dissipation becomes extremely large.
Figure\,4a shows that the magnetic structure in the 
galactic plane is not so aligned
with the bar as 
for the model Ib (Fig.\,3a, bottom). The magnetic arms are much broader
and big parts of them are completely dispersed. The main magnetic arm
which survives is that one with magnetic vectors directed opposite to the
initial direction of the magnetic field (Fig.\,4a, top) and is
certainly not connected with dynamical structure of the model galaxy.
For $9\cdot 10^{8}~$yr the magnetic intensity maxima are again more dissipated,
leaving only two arms with opposite directions of the vectors (Fig.\,4a,
bottom) and with similar intensity (Fig.\,4a, right) and rather
weak magnetic arm on the left-hand side of the figure.
Figure\,4b shows magnetic intensity (right) and magnetic field vectors (both
in the galactic plane)
(left) for $9.5\cdot 10^{8}~$yr. At this stage of evolution the
magnetic energy density is extremely low due to very strong diffusion.
Almost the
whole structure is dispersed, however in the upper part of the figure
there could be seen two magnetic arms with the field reversal.
The numerical experiment with the high value of
the diffusion coefficient results in much more smoothed magnetic structures
(Fig.\,4) than for the simulations Ib (Fig.\,3).
The intensity of regions, where
magnetic vectors are distributed randomly, decreases fast and causes
higher contrast between smoothed and random vector areas (Fig.\,4).
%%%%%%%%%%%%%%%%%%%%%%%%%%%%%%%%%%%%%%%%%%%%%%%%%%%%%%%%%%%%%%%%%%
%%%% Fig.5 %%%%%%%%%%%%%%%%%%%%%%%%%%%%%%%%%%%%%%%%%%%%%%%%%%%%%%%
%%%%%%%%%%%%%%%%%%%%%%%%%%%%%%%%%%%%%%%%%%%%%%%%%%%%%%%%%%%%%%%%%%
\begin{figure}[thbp]
\epsffile[185 275 427 517]{5461F5.ps}
\caption[]{{} The topolines of the velocity field intensity and
magnetic intensity as a gray plot at $t~=~9\cdot 10^{8}~$yr
for the model Ib with no magnetic diffusion.}
\end{figure}
%%%%%%%%%%%%%%%%%%%%%%%%%%%%%%%%%%%%%%%%%%%%%%%%%%%%%%%%%%%%%%%%%%
%%%%%%%%%%%%%%%%%%%%%%%%%%%%%%%%%%%%%%%%%%%%%%%%%%%%%%%%%%%%%%%%%%
%%%% Fig.6 %%%%%%%%%%%%%%%%%%%%%%%%%%%%%%%%%%%%%%%%%%%%%%%%%%%%%%%
%%%%%%%%%%%%%%%%%%%%%%%%%%%%%%%%%%%%%%%%%%%%%%%%%%%%%%%%%%%%%%%%%%
\begin{figure}[htbp]
\epsfysize=9cm
\epsffile[310 275 627 517]{5461F6a.ps}
\caption[]{{\rm a:} The topolines of the magnetic field component $B_{\rm z}$
at $t~=~9.0\cdot 10^{8}~$yr in the galactic plane
for the model Ib.}
\end{figure}\addtocounter{figure}{-1}
\begin{figure}[htpb]
\epsfysize=9cm
\epsffile[50 275 427 517]{5461F6b.ps}
\caption[]{{\rm b:} The same as in Fig.\,6a
but 350 pc above the galactic plane.}
\end{figure}\addtocounter{figure}{-1}
%%%%%%%%%%%%%%%%%%%%%%%%%%%%%%%%%%%%%%%%%%%%%%%%%%%%%%%%%%%%%%%%%%
%%%% Fig.6c%%%%%%%%%%%%%%%%%%%%%%%%%%%%%%%%%%%%%%%%%%%%%%%%%%%%%%%
%%%%%%%%%%%%%%%%%%%%%%%%%%%%%%%%%%%%%%%%%%%%%%%%%%%%%%%%%%%%%%%%%%
\begin{figure}[htpb]
\epsfysize=9cm
{\epsffile[185 275 427 517]{5461F6c.ps}}
\caption[]{{\rm c:} The same as in Fig.\,6a
but 350 pc under the galactic plane.}
\end{figure}
%%%%%%%%%%%%%%%%%%%%%%%%%%%%%%%%%%%%%%%%%%%%%%%%%%%%%%%%%%%%%%%%%%
%%%% Fig.7 %%%%%%%%%%%%%%%%%%%%%%%%%%%%%%%%%%%%%%%%%%%%%%%%%%%%%%%
%%%%%%%%%%%%%%%%%%%%%%%%%%%%%%%%%%%%%%%%%%%%%%%%%%%%%%%%%%%%%%%%%%
\begin{figure}[htpb]\rule{0.0pt}{1cm}
   \epsfysize=8.5cm
\epsffile[185 275 427 517]{5461F7.ps}
\caption[]{{}  Time evolution of the magnetic energy density $E$ normalised
to the initial energy density $E_{0}$ for the following models:
Ia and Ib with no
magnetic diffusion for 51 and 101 grid points, respectively, Ic with 3D
magnetic diffusion coefficient $10^{26}\,$cm$^{2}$/s, Id with planar diffusion
coefficient $10^{27}~{\rm cm}^{2}/{\rm s}$ and Ie 
with planar diffusion coefficient
$3.6 \cdot 10^{26}~{\rm cm}^{2}/{\rm s}$.}
\end{figure}

In order to compare the structure of the magnetic field with the
velocity field intensity Fig.\,5 presents the topolines of
the velocity field intensity over the magnetic intensity
gray-plots for the case without the diffusion (model Ib, $t=9 \cdot
10^{8}$~yr).
The basic velocity minimum is situated in the galactic center, then
two maxima are visible under and above the galactic bar.
The magnetic maxima connected with the bar coincide to some extent with the
velocity maxima, however the magnetic spirals are distributed
slightly outward from the galactic center. The spiral arms are not observed
in the topolines map, which means that velocity field gradients
connected with them are not high.

The magnetic configuration is examined not only for the galactic
plane but also for levels above and under the XY plane.  We have
found that the magnetic field forms the configuration depending on the z
coordinate.
In Fig.\,6a the
topolines of the $B_{\rm z}$ component for the galactic plane
(level=11) at time
step $9\cdot10^{8}~$yr
is presented for the experiment Ib.
In Fig.\,6b and c we present parallel cuts of these $B_{\rm z}$ component
350 pc above (level=4) and
under (level=17) the galactic plane.
Comparing Fig\,6a with
Fig.\,6b it is visible
that the  magnetic  field  disappears in certain
regions. Above the galactic plane the
situation is just opposite (Fig.\,6c). Regions vanishing
at level=4 now
are clearly visible, but a few next regions disappear. The edge-on-view
of the model galaxy would give different structures above and under
the galactic plane.

%%%%%%%%%%%%%%%%%%%%%%%%%%%%%%%%%%%%%%%%%%%%%%%%%%%%%%%%%%%%%%%%%%
%%%% Fig.8a%%%%%%%%%%%%%%%%%%%%%%%%%%%%%%%%%%%%%%%%%%%%%%%%%%%%%%%
%%%%%%%%%%%%%%%%%%%%%%%%%%%%%%%%%%%%%%%%%%%%%%%%%%%%%%%%%%%%%%%%%%
\begin{figure*}[hptp]
   \epsfysize=15cm
\epsffile[30 280 330 733]{5461F8a.ps}
\caption[]{\label{fig8a}{\bf a:}
Magnetic field vectors (left) and
intensity of magnetic field (gray plot, right) in the galactic plane
for the model IIa with no
magnetic diffusion at $t~=~3\cdot 10^{8}~$yr (top) and at
$t~=~4.5\cdot 10^{8}~$yr
 (bottom)}
\end{figure*}

The maps of the topolines of the $B_{\rm z}$ component (Fig.\,6) inform us
also that the magnetic field lines wave in
the Z~direction going along the spiral arms and bar.
The minima and maxima of the $B_{\rm z}$ component go one by one along these
structures changing their sign from plus to minus (Fig.\,6).
The amplitude of these waves is not
high - the value of $B_{\rm z}$ is of one order smaller than the value of
$B_{\rm x}$ and $B_{\rm y}$, but such waving is visible in most time
steps of the experiments with and without diffusion.

Figure\,7 presents the time evolution of magnetic energy density (E)
normalised to the initial energy density $E_{0}$ for the five
experiments which have been performed for the model I galaxy.
The figure legend is in agreement with
the description of the model I in Table 2. In order to solve the problem
of the numerical diffusion two experiments with no additional
magnetic dissipation have been made. In the case Ia two times larger
distance between grid points have been applied. The simulations Ib
result in significantly higher increase of magnetic energy density than
for the case with larger step. It means that numerical dissipation due
to truncation errors is higher for the bigger grid interval. We estimate
that in the case Ia the numerical diffusion is of order
$10^{26}$ cm$^2$/s and about one order less for the case Ib. For both
simulations the growth of magnetic energy density is present, however
the increase in the experiment Ib is certainly higher. The initial three maxima
of E are followed by quick decaying of its value due to
the effects of numerical diffusion. The magnetic energy evolution for
the model Ib could be compared with the case Ie with the same number of grid
points and grid interval. The simulations Ie have been done for the
planar coefficient of diffusion $\eta=3.6\cdot 10^{26}$ cm$^2$/s.
Figure\,7 shows that a small increase of magnetic energy density is still
present in this case, however after very short evolutionary time ($8
\cdot 10^{8}$~yr) a fast decreasing appears.
The models Ic and Id are prepared for 51 grid points in the X and Y
directions and could be compared with the case Ia. The simulations Ic and
Id have been done for three-dimensional diffusion $\eta=10^{26}$ cm$^{2}$/s
and for the planar dissipation $10^{27}$ cm$^2$/s, respectively. For
both cases the magnetic energy density E simply decreases with the
evolutionary time due to the numerical and physical diffusion. The
diminishing of E is faster for the computations Id with the biggest
magnitude of the diffusion coefficient. The decaying of E is connected
with  magnetic energy outflow through X and Y borders of the model
galaxy, while for the three-dimensional diffusion (Ic) outflow is present
mainly in the Z direction.

The maxima of the energy density evolution in case Ia and Ib can be
clearly identified as driven by the different stages of bar evolution.
At timestep 76 the bar starts to grow, reaching its maximum strength and length
at timestep 81 and then it shrinks slowly (up to timestep 90) followed
by an accelerated shrinkage.

All of our simulated models result in the decay of magnetic energy
density due to the numerical and/or physical diffusion. Each model
suggests the amplification of magnetic field in the arm due to
compressional and shearing motions of gas induced by the bar and arms.
The models with the diffusion working mainly in the XY direction (model
Ie) show much slower decline of the energy density.

\subsection{The 3D evolution of magnetic field: model II}

The calculations of magnetic field evolution with gravitationally
weak bar disturbance (model II) have been performed for earlier
time stages than in the model I.

%%%%%%%%%%%%%%%%%%%%%%%%%%%%%%%%%%%%%%%%%%%%%%%%%%%%%%%%%%%%%%%%%%
%%%% Fig.8b %%%%%%%%%%%%%%%%%%%%%%%%%%%%%%%%%%%%%%%%%%%%%%%%%%%%%%%
%%%%%%%%%%%%%%%%%%%%%%%%%%%%%%%%%%%%%%%%%%%%%%%%%%%%%%%%%%%%%%%%%%
\addtocounter{figure}{-1}
\begin{figure*}[hbtp]
   \epsfysize=7.5cm
\epsffile[20 490 400 740]{5461F8b.ps}
\caption[]{{\bf b:}
Magnetic field vectors (left) and
intensity of magnetic field (gray plot, right) in the galactic plane
for the model IIa with no
magnetic diffusion at  $t~=~5\cdot 10^{8}~$yr}
\end{figure*}

In order to illustrate high dynamics of the simulation process,
the evolution of stars and molecular clouds
is presented in (Fig.\,2a-c) for three evolutionary stages:
$t=2 \cdot 10^{8}~$yr, $t=3 \cdot 10^{8}~$yr and $t=5 \cdot
10^{8}~$yr. The magnetic field behaviour in this case is shown for
later moments of time (Fig.\,8).
Figure 8 presents the gray plots of the intensity and the vectors
of the magnetic field in the galactic plane, for the model IIa (see Table~2),
without the magnetic diffusion. Figure\,8a, top, corresponds 
to the evolutionary
stage of $t~=~3\cdot 10^{8}~$yr,
Fig.\,8a, bottom, to $t~=~4.5 \cdot 10^{8}~$yr,
and Fig.\,8b to $t~=~5\cdot 10^{8}~$yr. Already for the first time of
evolution, the initially
circular magnetic field configuration is transformed into a
structure with a bar and spiral arms. The direction of vectors is
mainly anti-clockwise, however in the inner region of the upper arm
(Fig.\,8a, top-left) the small area with vector reversal appears.
The magnetic intensity gray-plot (Fig.\,8a, top-right) shows nonuniform
distribution of the magnetic arms, which consist of two, three stripes
due to a similar behaviour of the dynamical arms (Fig.\,2b).
According to dynamical features, after
$1.5\cdot 10^{8}~$yr (Fig.\,8a, bottom) the magnetic arms
become more narrow. The intensity along the arm 
(Fig.\,8a, bottom-right) is still nonuniform
forming some kind of bifurcation regions. The magnetic field waves in the XY
plane in the lower magnetic arm (Fig.\,8a, bottom-left). The
region with opposite direction
of magnetic vectors inside the upper spiral arm is again visible,
however its intensity is rather weak (Fig.\,8a, bottom-right).
The magnetic field vector
mixing is present in both time steps in Fig.\,8a.
Figure 8b illustrates that magnetic field structure is different from the
molecular gas distribution (Fig.\,2c). In the lower part of the figure
the magnetic bridge between two arms is present. The intensity map
(Fig.\,8b, left) presents that the magnetic arms are nonuniform and wave
in the XY plane.

The magnetic time evolution of energy density E normalised 
to the initial energy
density $E_{0}$ is illustrated in Figure\,9.
The simulation for the model IIa results in the initial growing of E. After
two close maxima, E decreases with the evolutionary time due to numerical
dissipation. In the case IIb with 3D magnetic diffusion the magnetic
energy density simply diminishes with time due to physical as well as
numerical diffusion.

Calculations for model II result also in magnetic field mixing
and waving of the magnetic lines of force in the Z direction.

The magnetic field diminishing in the central part in all experiments is
connected with assumed lower magnetic strength in this region and with
numerical dissipation due to truncation errors.

\section{Conclusions and discussion}

The evolution of the large-scale galactic magnetic field affected
by non-axisymmetric dynamical structure like spirals and bars has
been demonstrated using 3D numerical simulations. The problem is
solved with three-dimensional, fully dynamical velocity field
evolving in time, resulting from N-body particle hydrodynamics,
as the input parameter. The calculations have been performed for
two dynamical experiments with strong bar and clearly seen spiral
structure at the end of a bar. The basic configuration for the
initial magnetic field is axisymmetric structure. The case of the
uniform input magnetic field has been demonstrated in Paper I
and missed as not interesting in the actual project.

During the first time steps of magnetic field evolution the lines
of force follow closely the non-axisymmetric dynamical structures.
After about $5\cdot 10^{7}~$yr the situation changes and the magnetic
fields become more complicated than the velocity field at
the given time step. The regions with magnetic vectors going opposite
to the main magnetic spirals are visible. This fact is connected
with the molecular gas flows in the dynamical structure.
Such reversals could probably explain the observed
reversals of magnetic field vectors in our Galaxy (Rand and Lyne 1994).
Further evolution shows more and more complicated structures of magnetic
lines of force and after the whole evolutionary time the magnetic
field is still aligned with the bar and spirals and the regions
with reversals are still present. In the simulation with
planar diffusion they are still seen after $9.5\cdot 10^{8}~$yr.
%%%%%%%%%%%%%%%%%%%%%%%%%%%%%%%%%%%%%%%%%%%%%%%%%%%%%%%%%%%%%%%%%%
%%%% Fig.9 %%%%%%%%%%%%%%%%%%%%%%%%%%%%%%%%%%%%%%%%%%%%%%%%%%%%%%%
%%%%%%%%%%%%%%%%%%%%%%%%%%%%%%%%%%%%%%%%%%%%%%%%%%%%%%%%%%%%%%%%%%
\begin{figure}[htbp]
   \epsfysize=8.5cm
   \epsffile[185 275 427 517]{5461F9.ps}
\caption[]{\label{fig9}
Time evolution of the magnetic energy density $E$ normalised
to the initial energy density $E_{0}$ for the model II with no
magnetic diffusion (dashed line) and  with 3D diffusion (solid line)}
\end{figure}

Moreover the magnetic field arms are not single features. They form
two or three stripes which change their shape during the evolution.
The magnetic intensity in arms is not uniform creating some kind of
bifurcations along the arms and bar. The presented simulations
resulted also in the waving in the XY plane of magnetic field vectors
(Fig.\,8b, left).

The regions with magnetic field
mixing are present at both ends of the bar. This fact results from
the rotation of the dynamical structure. The end of the  bar is the
place where the field aligned along two opposite sides meets
(see Figs. 3b, bottom \& 4b).
The magnetic field vectors have opposite directions going along two
bar sides and during rotation they meet at the bar ends.

Both obtained facts: magnetic lines alignment along the spiral arms
and mixing of the magnetic vectors at the bar ends could
explain the characteristic radio features observed in the barred
galaxy M83 (Neininger et al. 1991), however
the problem needs further study.
The model of polarised emission with the telescope beam and
the Faraday rotation for resulting magnetic field structures
is in preparation. The mechanism of magnetic field tangling
at the ends of a bar is normally explained by enhanced turbulent motions
of molecular clouds (e.g. Neininger et al. 1991).

The non-symmetrical effect of the levels from above and under the
galactic plane shows that the magnetised galactic disk is not uniform
along the Z-axis. The magnetic field changes its configuration from the
galactic plane to the z-edges of the disk. The magnetic field structure
is also highly different in the part above and under the galactic plane.
There is certainly no mirror symmetry according to this plane as it
is normally used in a number of simulations (e.g. Brandenburg et al. 1992).

The mechanism of magnetic field waving in the Z-axis constitutes
the next interesting result from our simulations. The $B_{\rm z}$ magnetic
component obtained from our calculations is of one order smaller than
$B_{\rm x}$ and $B_{\rm y}$. The magnetic field observations show that
the magnetic field in the direction perpendicular to the galactic
plane is not high (Dumke et al. 1995). Such behaviour of magnetic lines
of force has been found observationally in M31 (Beck et al., 1989) and probably
could help to initiate development of the Parker
instability in the spiral arms and bar (Hanasz and Lesch 1993).

Finally, we discuss the time-evolution of the magnetic energy density
$W_B=E/E_0$. We have already mentioned that the decline of $W_B$
is mainly related to the numerical diffusion which could be even of
$10^{26}~\rm cm^{2}/\rm s$ order in the case of big grid interval.

Our simulations clearly indicate the relation between the non-axisymmetric
dynamical perturbations and the magnetic field evolution as long as the
diffusion is not too strong. Any local deviation
from the global velocity field is immediately visible in the magnetic field
strength and structure. We emphasize that
in case Ia and Ib
the different stages of bar dynamics can be found in the evolution of
the magnetic field energy density $W_B$.
Growth of a bar up to its maximum strength (time step 76 to 81)
and its subsequent shrinkage 
(slow up to time step 90 faster for time steps $> 90$) )
can be easily related with the maxima of the energy density time profile.
The same holds for model II.
After a first maximum is reached ($3.2\cdot
10^8$ yr), which
is correlated with the appearance of the 2 arm spirals, $W_B$
drops a little and then increases again. The second maximum
appears when the bar reaches his maximum length and strength
($3.9\cdot 10^8$ yr). After $t=4 \cdot 10^{8}~$yr
$W_B$ decreases due to numerical diffusion connected
with the truncation errors.

%%%%%%%%%%%%%%%%%%%%%%%%%%%%%%%%%%%%%%%%%%%%%%%%%%%%%%%%%%%%%%%%%%
%%%% Fig.10 %%%%%%%%%%%%%%%%%%%%%%%%%%%%%%%%%%%%%%%%%%%%%%%%%%%%%%%
%%%%%%%%%%%%%%%%%%%%%%%%%%%%%%%%%%%%%%%%%%%%%%%%%%%%%%%%%%%%%%%%%%
  \begin{figure*}[htpb]
   \epsfysize=5.5cm
    \rotate[r]{\epsffile[10 10 370 290]{5461F10.ps}}
\caption[]{\label{sigfig}Comparison between the two bar models. The left panel
shows the surface density $\Sigma$, the right panel
the rotation velocity $v_{\phi}$ of two models}
\end{figure*}

Obviously, the magnetic field reacts sensitively to variations
in the gas velocity, triggered by dynamical processes. These
variations can be seen from Fig.\,10\,a and b where we plotted the surface
mass density $\Sigma$ and the gas rotation curves for both
models at time steps where in both models a bar has finished its growth.
The differences in the dynamical behaviour of both models are
obvious. In model II there is much more structure in the rotation
curve and a higher central mass concentration. Dynamically speaking
 model II is more ``active''.

This enhanced activity is easy to understand, since the gas constant of the
simulated galaxy is higher in model II. Dynamical disk instabilities
rely on the ratio of gas mass to total mass, and a higher gas contents
leads to a stronger dynamical disk activity.

\acknowledgements{
We thank Prof R. Wielebinski for financial support and
the hospitality of Max-Planck-Institute f\"ur
Radioastronomie in Bonn.
SvL thanks Prof. F. Combes for letting us her code.
KO wishes also to express her gratitude to Marek Urbanik \& Marian Soida for
their valuable advices during this work.
Calculations have
been supported by the Forschungszentrum J\"ulich GmbH and
ACK CYFRONET-KRAK\'OW.
This work was partly supported
by the grant from Polish Committee for Scientific Research (KBN),
grant no. PB/0578/P03/95/09.}

%%%%%%%%%%%%%%%%%%%%%%%%%%%%%%%%%%%%%%%%%%%%%%%%%%%%%%%%%%%%%%%%%%%%%%%%%%%%
%%%%%%%%%%%%%%%%%%%%%%%%%%%%%%%%%%%%%%%%%%%%%%%%%%%%%%%%%%%%%%%%%%%%%%%%%%%%
%%%%%References
%%%%%%%%%%%%%%%%%%%%%%%%%%%%%%%%%%%%%%%%%%%%%%%%%%%%%%%%%%%%%%%%%%%%%%%%%%%%
%%%%%%%%%%%%%%%%%%%%%%%%%%%%%%%%%%%%%%%%%%%%%%%%%%%%%%%%%%%%%%%%%%%%%%%%%%%%

\end{document}